\documentstyle[12pt]{article}

\newcommand{\EQ}{\begin{equation}}
\newcommand{\EN}{\end{equation}}
\newcommand{\bea}{\begin{eqnarray}}
\newcommand{\ena}{\end{eqnarray}}
\newcommand{\bdis}{\begin{displaymath}}
\newcommand{\edis}{\end{displaymath}}
\newcommand{\vs}[1]{\vspace{#1 mm}}

\renewcommand{\a}{\alpha}
\renewcommand{\b}{\beta}
\renewcommand{\c}{\gamma}
\renewcommand{\d}{\delta}
\renewcommand{\o}{\omega}
\renewcommand{\t}{\tau}
\newcommand{\tq}{\tilde q}

\newcommand{\pa}{\partial}

\newcommand{\nn}{\nonumber \\}

\begin{document}

\topmargin 0pt
\oddsidemargin 5mm

\newcommand{\NP}[1]{Nucl.\ Phys.\ {\bf #1}}
\newcommand{\PL}[1]{Phys.\ Lett.\ {\bf #1}}
\newcommand{\CMP}[1]{Comm.\ Math.\ Phys.\ {\bf #1}}
\newcommand{\PR}[1]{Phys.\ Rev.\ {\bf #1}}
\newcommand{\PRL}[1]{Phys.\ Rev.\ Lett.\ {\bf #1}}
\newcommand{\PTP}[1]{Prog.\ Theor.\ Phys.\ {\bf #1}}
\newcommand{\PTPS}[1]{Prog.\ Theor.\ Phys.\ Suppl.\ {\bf #1}}
\newcommand{\MPL}[1]{Mod.\ Phys.\ Lett.\ {\bf #1}}
\newcommand{\IJMP}[1]{Int.\ Jour.\ Mod.\ Phys.\ {\bf #1}}
\newcommand{\AP}[1]{Ann.\ Phys.\ {\bf #1}}

\begin{titlepage}
\setcounter{page}{0}
\begin{flushright}
NBI-HE-94-19\\
March 1994\\
\end{flushright}

\vs{8}
\begin{center}
{\Large Improved Wick Rotation Prescription in \\
Stochastic Quantization of Dissipative Systems}

\vs{15}
{\large Naohito Nakazawa\footnote{e-mail address:
 nakazawa@nbivax.nbi.dk, nakazawa@ps1.yukawa.kyoto-u.ac.jp}$^,$\footnote{
Permanent address: Department of Physics, Shimane University,
Matsue 690, Japan}}\\
{\em The Niels Bohr Institute, Blegdamsvej 17, DK-2100 Copenhagen \O,
Denmark}

\vs{8}
{\large Eisaku Sakane}\\
{\em Department of Physics, Shimane University,
Matsue 690, Japan}
\end{center}

\vs{8}
\centerline{{\bf{Abstract}}}

We apply Stochastic Quantization Method to
dissipative systems at finite temperature. Especially, the relation
of SQM to the Caldeira-Leggett model is clarified ensuring that
the naive Wick rotation is improved in this context.
We show that the Langevin system obtained by
the \lq\lq Improved Wick Rotation " prescription is
equivalent to an ideal friction case ( low temperature limit)
in the C-L model. We derive, based on our approach, a general formula
on the fluctuation-dissipation theorem for higher derivative frictions.

\end{titlepage}
\newpage
\renewcommand{\thefootnote}{\arabic{footnote}}
\setcounter{footnote}{0}

Quantization of dissipative
systems has been discussed especially in connection with the description of
non-equilibrium quantum phenomena.
To describe the quantum effect in
dissipative systems, one may naturally consider the dissipative
system as a part of a \lq\lq  whole system ".
In a physical point of view, the
macroscopic dissipation
comes from the microscopic interaction between the system
and its environment.
Realizing this idea explicitly, Cardeila and Leggett (C-L)
developed a model to describe dissipation~\cite{CL}.
They introduced
artificially harmonic oscillators as the environment,
which couples with the system in question
linearly. By integrating out the environmental degrees of freedom,
they derived an effective action for the quantum effects with
dissipation.

In the previous paper~\cite{NSU},
we applied Stochastic Quantization Method (SQM)~\cite{PW} on the
description of the quantum effect with dissipation.
Since SQM can be formulated without action principle, namely it is
formulated in terms of a Langevin equation which includes
an equation of motion as the drift force,
the artificial introduction of the environmental degrees of
freedom is not necessary for the description of dissipative systems.
This point may be an advantage of SQM in formulating
a quantization of dissipative systems because it avoids
the possible dependence of the physical results
on the choice of the artificial environment
and its interaction with the system in question.
Our basic motivation is to clarify, whether SQM
can describe a system interacting with its environment which causes
dissipation by starting from the phenomenological equation of motion
including a dissipative term.
In Ref.~\cite{NSU}, we considered the application of SQM to
quantum mechanical systems with friction.
We showed that some improvements are necessary
to describe dissipative systems in SQM. As a guiding principle,
we took our attention especially on the correspondence of
our approach to the fluctuation-dissipation theorem and proposed an improved
Langevin equation to describe the quantum mechanical systems with dissipation
at finite temperature. In this paper, we show the equivalence
of the SQM approach in terms of the improved Langevin
equation to the C-L model approach at finite temperature.

We first recapitulate the essence of Ref.[2]. Let us
consider the system with dissipation for which the equation of motion is
given by,
\EQ
M{\ddot q} + {\pa \over\pa q}V(q) + \eta {\dot q} = 0 .
\EN
Here $\eta$ is a phenomenological friction coefficient. The dot denotes the
derivative with respect to the real time coordinate $t$. In a naive sense,
we may start with a Langevin equation\footnote{
For the
equation of motion (1), one can write
a time dependent lagrangian. Then, we would start from the explicitly
time dependent
Langevin equation,
\bdis
{d\over d\t}q(\t,t) =
- i f(t) \big\{ M{\ddot q} + {\pa \over\pa q}V(q) + \eta {\dot q}
\big\} + \xi (\t,t)  ,
\edis
with $f(t) = {\rm e}^{(\eta/M)t}$.
In this case, however,
the explicit time dependence would cause some difficulties.
It is also shown that the approach in terms of the above Langevin equation
is, in a rigorous sense, not equivalent to the C-L approach\cite{NSU}.
},
\EQ
{d\over d\t}q(\t,t) =
- i \big\{ M{\ddot q} + {\pa \over\pa q}V(q) + \eta {\dot q} \big\}
+ \xi (\t,t)  ,
\EN
where the correlation of the white noise $\xi(\t, t)$ is defined by
\EQ
<\xi(\t,t) \xi(\t',t')>
= 2\hbar \d (\t - \t') \d (t - t')   .
\EN
In the systems without dissipation, (2) with $\eta = 0$,
the finite temperature Langevin equation is derived in a heuristic way as
follows.                \\
(a): Continue the real ( Minkowski ) time to the imaginary ( Euclidean )
time by Wick
rotation; $t \rightarrow -i t$.               \\
(b): To improve the Langevin equation such that
it has an equilibrium limit, \lq\lq Wick " rotate the fictitious time
$\t$ into $-i\t$. ( This procedure makes the drift force
positive definite in
the Langevin equation.)    \\
(c): Redefine the white
noise variable $\xi \rightarrow i\xi$ to preserve the noise correlation (3)
invariant
under these continuations, (a) and (b).            \\
(d): Require that
all the variables satisfy the periodic ( anti-periodic ) boundary condition
for bosonic system ( fermionic system ).            \\
The procedure (a)-(d) really recovers the imaginary time method at finite
temperature\cite{GN}\cite{Na}.
Following to this naive \lq\lq Wick rotation " prescription,
we found that we couldn't obtain the Langevin equation which
describes the damped oscillator system at finite temperature.
As it is clarified in Ref.[2], the failure comes from the retarded
( causal ) nature of the Green's function obtained at zero temperature.
Then we proposed the following improved
Langevin equation for a damped oscillator case.
\bea
{d\over d\t}\tq_n(\t)
&=&  - M(\o_n^2 + \o_0^2 ) \tq_n(\t) - \eta | \o_n |
\tq_n(\t)
+ {\tilde \xi}_n (\t)      ,               \nn
<{\tilde \xi}_m(\t){\tilde \xi}_n(\t')>
&=& 2\hbar \d (\t - \t')\b\hbar\d_{m+n,0}   .  \nn
\ena
The expectation value, $<{}>$, means the finite
temperature expectation value.
The tilde denotes its Fourier component\footnote{
The coefficients of the Fourier series are defined by
\bdis
q(\t, t)
= \sum_n {1\over\b\hbar}\tq_n(\t){\rm e}^{-i\o_n t}       ,
\edis
with $\o_n = {2\pi n\over \b\hbar}$. $t$ denotes the imaginary time.
}.
Here we have taken the additional procedure    \\
(e): Take the absolute value of the Fourier momentum in the
friction dependent term in the momentum
space Langevin equation.  \\
It was shown that the momentum space Langevin equation
(4) really recovers the
result which is consistent to the fluctuation-dissipation theorem.
As it can be seen from (4),
the absolute value prescription has been taken to make
the drift force positive definite.
We notice that the absolute value $| \o_n |$ would not
appear in the right hand side but $\o_n $ would appear if we would
consider the naive Wick rotation (a)-(d).
Thus the absolute value prescription (e)
for the friction dependent term
is necessary for the existence of an equilibrium distribution
at the infinite fictitious time limit.
For a general potential, $V(q)$,
one can obtain the Langevin equation by simply replacing
\EQ
M\o_0^2 \tq_n(\t) \rightarrow
\int^{\b\hbar}_0 \! dt {\rm e}^{i\o_n t} {\pa \over \pa q}
V\big( v + q(\t, t)\big) ,
\EN
in the r.h.s. of (4). Here the
vacuum $v$ satisfies ${\pa\over \pa q}V(v)=0$.

Before showing the equivalence of our approach to the C-L model approach,
in demonstrating how our approach works well, we consider the case of
higher derivative frictions which is defined by the following
phenomenological equation of motion.
\EQ
M{\ddot q} + {\pa \over\pa q}V(q) + (-)^{l-1}\eta q^{(2l-1)} = F_{ex} ,
\EN
where
$
q^{(2l-1)} \equiv d^{2l-1}q(t)/dt^{2l-1} .
$
The external force $F_{ex}(t)$ is introduced for the later
convenience. $l=1$ corresponds to
Ohmic dissipation in (1) and $l=2$ corresponds to super-Ohmic case\cite{FISS}.
The sign factor, $(-)^{l-1}$, comes from that fact that the system is
dissipative. Actually, the average of
the energy,
$
{1\over2}M{\dot q}^2 + V(q)
$
, in a periodic motion is dissipated with the rate
$
\eta \{ d^l q(t)/d^l t \}^2 .
$
Thanks to the improved Wick rotation prescription (a)-(e),
we may begin with the
Langevin equation
\bea
{d\over d\t}\tq_n(\t)
&=&  - M(\o_n^2 + \o_0^2 ) \tq_n(\t) - (-)^{l-1}\eta | \o_n |^{2l-1}
\tq_n(\t)
+ {\tilde \xi}_n (\t)      ,               \nn
<{\tilde \xi}_m(\t){\tilde \xi}_n(\t')>
&=& 2\hbar \d (\t - \t')\b\hbar\d_{m+n,0}   .  \nn
\ena
for $F_{ex}=0$. The factor $(-)^{l-1}$ comes from the same factor
appears in (6).
For simplicity, we mainly consider the harmonic oscillator
with $l = odd$ case, however,
the generality is obviously recovered by the replacement (5) in (7).
By using the solution of (7), we obtain the two-point
function at finite temperature
\bea
<{q^2}>
&\equiv&  \lim_{\t = \t' \rightarrow \infty}
<{q(\t,t)q(\tau',t)}>         ,\nn
&=& \sum_n {1\over\b} \Bigl\{ {1\over {M(\o_n^2 + \o_0^2)
+ (-)^{l-1}\eta |\o_n|^{2l-1}}} \Bigr\}, \nn
\ena
By using the same technique in Ref.[1], we obtain an equivalent
expression to (8) as follows,
\bea
<{q^2}>
&=& {\hbar\over\pi}\int^\infty_0  {\rm coth}({\b\hbar\o\over2})
{\rm Im}\chi_q (\o) d\o           \nn
&-& {\hbar\over\pi}
\sum_{\a_p \in {\rm u.h.p.}}{\rm Im} \int_{\a_p}
\big\{ \chi_q (\o) {1\over {\rm e}^{\b\hbar\o} - 1 } \big\} d\o    ,\nn
\chi_q (\o)
&\equiv& ({\d q \over \d F_{ex}})_\o   ,\nn
&=& {1\over M( \o_0^2 - \o^2 - 2i \c \o^{2l-1})}    ,
\quad \c \equiv  \eta /2M          ,\nn
\ena
where $\a_p$ denotes the pole of $\chi_q(\o)$ in the complex $\o$-plane.
The summation $\sum_{\a_p \in {\rm u.h.p.}}$ means that the only poles on
the upper half $\o$-plane should be summed up. For the case of $l=1$,
Ohmic dissipation, the formula turns out to be equivalent to the
result appeared in Ref.[1]\footnote{
In the Ohmic dissipation, l=1, the second term is not necessary because
the poles appears only on the lower half of the complex $\o$-plane.}
. We notice that the second term in the r.h.s. of
the first relation in (9) is not
a correction on the fluctuation-dissipation theorem which connects
$<q^2>$ to the imaginary part of the response function
, ${\rm Im}\chi_q$, but a precise reflection of the statement
of this theorem that
the contributions which come from only the retarded Green's function
defined by the equation of motion (6) should be included in
this formula. If the second term would not appear in the r.h.s. of
(9), the causal structure of this formula would be violated.
In this sense, the second term in (9) is recognized as the counter term
for the contribution from u.h.p.,
possibly included in ${\rm Im}\chi_q$, which may
violate the causal structure of this formula.

%
%

We remark that the Fokker-Plank equation which corresponds to (7) with
(5) gives
the following equilibrium distribution functional in momentum space at
the infinite fictitious time limit,
\EQ
P_{\rm eq}[q]=N\exp\Bigl[ -{1\over\hbar}
\sum_n{1\over{2\b\hbar}} \Bigl\{
M\o_n^2 + (-)^{l-1}\eta|\o_n|^{2l-1} \Bigr\} |\tq_n|^2
-{1\over\hbar}\int^{\b\hbar}_0 \! dt {\rm e}^{i\o_n t}
V\big( v + q(\t, t)\big)      \Bigr] ,
\EN
provided that the potential ensures the positive
definite drift force in (7) such as in the double well
potential case\cite{FISS}.
For the harmonic oscillator with $l = even $,
the Langevin equation (7) does not have any equilibrium
limits. This is a difficulty not only for SQM
approach but also for the corresponding C-L model approach because, as
we will discuss later, the path-integral does not converge in this case.

The remaining part of this paper is devoted to show
the equivalence of the prescription (e)
to the microscopic description in terms of C-L model.
Now we clarify the relation of (7) with (5)
to C-L model at finite temperature.
The effective action of C-L model at finite temperature is given by
\cite{CL}\footnote{
In this effective action,
the necessary subtraction of the \lq\lq divergent " term to
recover the phenomenological equation of motion such as (1) and (6) is
included\cite{CL}\cite{FISS}\cite{CK}.
Thus the potential appears in (11) is recognized as
the renormalized one. The reader desiring a full account should consult
Ref.\cite{FISS}.}
%
\EQ
S_{\rm eff} = \int^{\b\hbar}_{0}\! d\t\big[
{M\over 2}{\dot q}^2 + V(q) \big] +
\int^{\b\hbar}_{0}\! dt \!\int^{\b\hbar}_{0}\! dt' K(t - t')
q(t)q(t')       ,
\EN
We notice that the dot denotes the derivative with respect to
the imaginary time $t$ in the following arguments.
The Fourier components of the kernel
function $K(t)$ is defined by\cite{CL}\cite{CK}
\EQ
K(\o_n) = {\o_n^2 \over \pi}\int^\infty_0\! d\o {1\over\o}
{J(\o)\over {\o^2 + \o_n^2}}       .
\EN
The explicit form of the spectral function $J(\o)$, which is determined by
requiring that the phenomenological equation of motion (1) should be
recovered in a classical limit,
is $not$ crucial in the
following arguments.
This is because, from (12),
the kernel function $K(\o_n)$ is an even function of $\o_n$,
a function of $|\o_n|$, ensuring the validity of the
absolute value prescription (e).
For the Ohmic dissipation, it is required to be
$
J(\o) = \eta\o     .
$
While in a physically realistic situation, the dependence of $J(\o)$ on
the high frequency $\o$ would be modified depending
on its microscopic physics. We here only assume,
\EQ
J(\o) = \eta \o {\rm e}^{- \o /\o_c} ,
\EN
by introducing a critical value $\o_c$\cite{CK}.
Then $K(\o_n)$ is given by
\EQ
K(\o_n) = {\eta\over2} {|\o_n|\over 1 + |\o_n|/\o_c}.
\EN
Now the correspondence of our approach and C-L model is obvious. At
the limit $\o_c \rightarrow \infty$, the
Fokker-Planck
distribution (10) is equivalent to path-integral measure via
the effective action of C-L model.
Namely, in this limit,
(4) is equivalent to the Langevin equation derived from the
effective action of C-L model at finite temperature,
\EQ
{d\over d\t}q(\t,t) =
- {\delta S_{\rm eff}\over\delta q}(\t, t) + \xi (\t,t) .
\EN
Actually, we have the following Langevin equation as
the configuration space representation of (4).\footnote{
We also find another useful form of the configuration
space expression of (4),
\bdis
{\pa\over\pa\t}q(\t,t)
=M\ddot q
-{\pa V\over\pa q}
-{\eta\over\pi}
\int_{-\infty}^\infty dt' {{q(\t, t)-q(\t, t')}\over(t-t')^2}+\xi(\t,t) .
\edis
Here we $define$ the integral by its principal value.
}.
%
\EQ
{\pa\over\pa\t}q(\t,t)
=M\ddot q
-{\pa V\over\pa q}
-{\eta\over\pi}
\int_{0}^{\b\hbar}\!dt' {{q(\t, t)-q(\t, t')}\over
\{ ({\b\hbar\over\pi}){\rm sin}[\pi (t -t')/\b\hbar]  \}^2
}+\xi(\t,t) .
\EN
%

This correspondence is easily extended to the higher derivative
friction cases.
For the higher derivative case, to recover the phenomenological
equation of motion (6), the spectral function is assumed to be
\EQ
J(\o) = \eta \o^{2l-1} {\rm e}^{- \o /\o_c} ,
\EN
Then we have the kernel function $K(\o_n)$ in the limit
$\o_c \rightarrow \infty$,
\EQ
\lim_{\o_c \rightarrow \infty} K(\o_n)
= (-)^{l-1}{\eta\over2}|\o_n|^{2l-1}   .
\EN
This ensures the complete equivalence of the Fokker-Planck distribution
(10) to the path-integral measure via C-L model effective action.

In a physically interesting case, $l=2$ ( super Ohmic case ), the explicit
form of the Langevin equation in configuration space is given by
\EQ
{\pa\over\pa\t}q(\t,t)
=M\ddot q
-{\pa V\over\pa q}
- {\eta\over\pi}{d\over dt}\Bigg[
\int_{0}^{\b\hbar}\!dt' {{{\dot q}(\t, t)-{\dot q}(\t, t')}\over
\{ ({\b\hbar\over\pi}){\rm sin}[\pi (t -t')/\b\hbar]  \}^2} \Bigg]
+\xi(\t,t) .
\EN
The friction dependent term in the r.h.s is the variation of the effective
action of C-L model in super Ohmic case\cite{FISS}.

It is clear from (18) that the Langevin equation for the higher derivative
friction case can not be obtained by the naive
Wick rotation prescription from the
friction dependent term $\eta d^{2l-1}q(t)/dt^{2l-1}$.
We have thus confirmed the validity of the absolute value prescription (e)
in these cases as well.
We notice, however,
the appearance of the factor $(-)^{l-1}$ in
the phenomenological equation of motion (6),  which
comes from the positivity of the
spectral function $J(\o)$ as is clearly
recognized in C-L model approach, would cause a difficulty
not only in SQM approach but also in C-L model approach
\footnote{
For the super Ohmic case, the opposite sign factor appears, as it is
clear from (7), in the friction dependent term which comes from (6).
In this case, if the friction coefficient $\eta$
would exceed its critical value
which makes the friction term comparable to the \lq\lq kinetic +
potential " term, we
could not obtain the equilibrium limit in the Langevin equation
(7) with (5).
}.
In some cases with $l = even$, for example a harmonic oscillator,
the path-integral does not converge with
the effective action (11).
The best way to derive an effective action in C-L model which
avoids this difficulty would be to introduce a
cut off in the high frequency part of the spectral function such as,
\EQ
J(\o) = \eta \o^{2l-1}\theta ( \o - \o_c ) ,
\EN
where $\o_c$ is a critical value of the frequency which is determined
by the requirement for the effective action (11) to be bounded from the below.
This would yield a modification on the
phenomenological equation of motion such as (6) ensuring a positive
definite drift force in the improved Langevin equation. It is,
however, beyond C-L model approach as well as SQM approach to
justify such an ad hoc modification (20). It depends on the detail of the
microscopic physics and we do not discuss on this possibility further.
In any cases,
what we would like to remark is that a well-defined path-integral measure
with the C-L model effective action which is bounded from the below
corresponds to an improved
Wick rotation prescription in SQM via the kernel, $K(\o_n)$,
the possible explicit form of which
is fixed as a function of $|\o_n|$ by dimensional analysis up to a
numerical factor.

In this short note, we have clarified the equivalence of
the \lq\lq improved Wick rotation "
prescription in SQM at finite temperature to C-L model approach.
Our main claim is that, to derive the Langevin equation by starting
from the phenomenological equation of motion, the naive Wick rotation
should be improved. We also have derived a generalized formula corresponding
to the fluctuation-dissipation theorem in higher derivative
friction cases.
We have mainly studied the limit that the low frequency dependence
of the spectral function is extrapolated to the high frequency region
($\o_c \rightarrow \infty$ in (13) and (17)).
In a physical point of view, this assumption
corresponds to the low temperature case,
$
{1\over \b\hbar} << \o_c  .
$
Thus it is more precise to conclude that
the improved Wick rotation prescription in SQM in quantizing
dissipative systems has its validity at low temperature.

\vs{10}
\noindent
{\it Acknowledgements}

The author (N.N.) wishes to thank all members at the Niels Bohr Institute
for warm hospitality.
This work is supported in part by the Grant-in-Aid, Ministry of Education.

\end{document}